\documentclass[a4paper,11pt]{article}

\usepackage{graphicx}
\usepackage{amsmath}
\usepackage{amssymb}
\usepackage{authblk}
\usepackage{fancyhdr}
\usepackage[titletoc]{appendix}
 
\usepackage{bm}
\usepackage{amsfonts}
\usepackage{epstopdf}
\usepackage{xcolor}
\usepackage{color}
\usepackage{wasysym}

\title{\sffamily{Highly collimated electron acceleration\\by longitudinal laser fields in a hollow-core target}}

\author[1,2]{Z. Gong}
\author[3]{A. P. L. Robinson}
\author[1]{X. Q. Yan}
\author[4]{A. V. Arefiev}

\affil[1]{State Key Laboratory of Nuclear Physics and Technology, and Key Laboratory of HEDP of the Ministry of Education, CAPT, Peking University, Beijing 100871, China}

\affil[2]{Center for High Energy Density Science, The University of Texas, Austin, TX 78712, USA}

\affil[3]{Central Laser Facility, STFC Rutherford-Appleton Laboratory, Didcot OX11 0QX, UK}

\affil[4]{Department of Mechanical and Aerospace Engineering, University of California at San Diego, La Jolla, CA 92093, USA}

\date{\today}

\begin{document}

\vskip -2.0cm
\maketitle
\vskip -3.0cm

\begin{abstract}
The substantial angular divergence of electron beams produced by direct laser acceleration is often considered as an inherent negative feature of the mechanism. The divergence however arises primarily because the standard approach relies on transverse electron oscillations and their interplay with the transverse electric fields of the laser pulse. We propose a conceptually different approach to direct laser acceleration that leverages longitudinal laser electric fields that are present in a tightly focused laser beam. A structured hollow-core target is used to enhance the longitudinal fields and maintain them over a distance much longer than the Rayleigh length by guiding the laser pulse. Electrons are injected by the transverse laser electric field into the channel and then they are accelerated forward by the pulse, creating an electron current. The forces from electric and magnetic fields of this electron population compensate each other, creating a favorable configuration without a strong restoring force. We use two-dimensional particle-in-cell simulations to demonstrate that a low divergence energetic electron beam with an opening angle of less than 5$^\circ$ can be generated in this configuration. Most of the energy is transferred to the electrons by the longitudinal laser electric field and, given a sufficient acceleration distance, super-ponderomotive energies can be realized without sacrificing the collimation.
\end{abstract}

\section{\sffamily{Introduction}}

The development of relativistically intense laser pulses~\cite{1985_CPA,ELI,XCELS} has enabled several schemes for producing laser-driven relativistic electron bunches.  One such scheme is Direct Laser Acceleration (DLA)~\cite{pukhov1999_DLA, gahn1999_DLA}. The normalized amplitude of the electric field in a laser pulse with wavelength $\lambda$ and intensity $I_0$ is defined as $a_0 \equiv \sqrt{I_0[W/cm^2] / 1.37\times10^{18}} \lambda [{\mu}m]$.  If $I_0$ exceeds $1.37\times 10^{18}$ W/cm$^2$, then this implies that the work done by the laser field during one period is comparable with the electron rest energy. As the electron velocity is boosted close to the speed of light in one laser cycle, the electron also gains considerable momentum in the direction of laser propagation via the Lorentz force, i.e. the ${\bf v}\times{\bf B}$ term in the equation of motion. If the electron were to maintain its relative phase with respect to the laser field then this longitudinal momentum $p_{\parallel}$ would become very large~\cite{gibbon2004short}. However, even in the case of an infinite plane wave, the natural relativistic dephasing rate limits this to $p_{\parallel}/m_e c = a_0^2/2$, where $m_e$ is the electron mass and $c$ is the speed of light. In reality, the momentum may be even more limited due to diffraction of the laser pulse~\cite{svelto1998_laser_principles, stupakov2001ponderomotive, robinson2018interaction}, which raises questions about the limits of what can be achieved with DLA.

A conventional approach to DLA is to employ a uniform plasma slab that not only restrains the laser defocusing~\cite{max1974self, wang2011laser} but also maintains a plasma channel with a positively charged and slowly evolving ion background~\cite{pukhov1999_DLA, arefiev2016beyond}. The ions generate a transverse electric field that confines laser-accelerated electrons inside the channel, preventing them from being prematurely expelled by the transverse ponderomotive force of the laser pulse. The ion electric field forces the accelerated electrons towards the axis of the channel and the ensuing transverse oscillations can be successfully leveraged to enhance the electron energy gain from the laser pulse~\cite{pukhov2002strong, arefiev2012_PRL, Khudik_POP_2016}. Even though a dense energetic electron bunch can be generated using this approach~\cite{liu2013_PRL, hu2015dense}, the bunch unavoidably has a relatively large divergence angle that can exceed $20^\circ$ due to the transverse oscillations. Such significant and difficult to control divergence might be undesirable for some potential applications. For example, the photon emission by the DLA electron bunch~\cite{ji2014radiation, ji2014energy,liu2015pop, Stark2016_PRL, chang2017brilliant, zgong_PPCF_2018} inherits the angular divergence which then impacts the brilliance of the resulting advanced light sources.

It has been recently shown that using structured targets can have a significant benefit if the end-goal is to generate a large number of energetic electrons in the form of a collimated beam~\cite{jiang2014effects, jiang2016microengineering}. For example, a microwire array attached to a solid density target can be used to guide the irradiating laser pulse and to provide a reliable source of electrons that can then be accelerated to high energies. In the case of a flat solid-density target, the electrons usually come from a pre-plasma whose characteristics are determined by the laser pre-pulse. As a result, experimentally observed electron acceleration is difficult to control and to predict using numerical simulations. A structured target, as the one considered in Refs.~\cite{jiang2014effects} and~\cite{jiang2016microengineering}, provides the much needed control over the electron acceleration.

Here, we continue the research of structured laser-irradiated targets by considering a hollow-core target whose channel width is smaller than the width of the irradiating laser beam. We report a new scheme for achieving a collimated laser-accelerated energetic electron bunch that utilizes longitudinal laser electric fields. By focusing a relativistically intense laser pulse into a hollow-core target, an electron channel, only filled with negative charge, can be produced in the hollow region. In contrast to the fields in the ion channel~\cite{arefiev2016beyond}, quasi-static electric and magnetic fields induce transverse forces that nearly compensate each other for forward-moving ultra-relativistic electrons. This provides an avenue to eliminate the transverse oscillations and reduce the angular divergence of the accelerated electrons. Moreover, the reduction of the transverse oscillations greatly facilitates electron acceleration by the longitudinal laser electric field, since the electrons can now remain in the favorable phase of the wave for much longer. Combined with the strengthened longitudinal laser field, a collimated longitudinal laser field acceleration is realized. 
 
The paper is organized as follows: In Sections~\ref{Sec-2} and \ref{Sec-3}, we show how the hollow-core target can produce the favorable channel fields. In Section~\ref{Sec-4}, the acceleration mechanism is demonstrated by tracking the accelerated electrons and the work performed by transverse and longitudinal laser electric fields. In Section.~\ref{Sec-5}, we show that the electrons accelerated by the longitudinal field can be highly collimated and that their energy can significantly exceed what is achievable in a pure vacuum case. In Section~\ref{Sec-6}, the key conclusions of this work are summarized. The details of the setup used in the simulations are given in the Appendix.


\section{\sffamily{Enhancement of the longitudinal laser field by a hollow channel}} \label{Sec-2}

Efficient acceleration via DLA requires that the laser pulse propagates in a stable fashion for a sufficiently long distance. Since we must, in reality, utilize focused laser pulses, the defocussing of the laser pulse, via diffraction, puts a fundamental limit on the longitudinal spatial range over which a given intensity can be maintained.  In the case of a Gaussian beam, this length is given by the Rayleigh length $l_R = \pi \sigma_0^2/\lambda$, where $\sigma_0$ is the beam waist in the focal plane and $\lambda$ is the laser wavelength. The transverse electric field of a beam propagating along the $x$-axis scales as 
\begin{equation}
	E_{\perp}^{3D} \propto \frac{\sigma_0}{\sigma(x)} \exp \left( - \frac{r^2}{\sigma^2(x)} \right),
\end{equation}
where $r$ is the radial distance from the beam axis and 
\begin{equation} \label{EQ-sigma}
\sigma(x) = \sigma_0 \sqrt{1 + \left. \left( x-x_0 \right)^2 \right/ l_R^2 }.
\end{equation}
The focal plane is located at $x = x_0$, so the beam radius is increased by a factor of $\sqrt {2}$ and the corresponding beam cross sectional area by a factor of 2 at $x = x_0 + l_R$. Therefore, the beam on-axis intensity at a Rayleigh length away from the beam waist is less by a factor of 2 than the peak intensity at the beam waist. If the Gaussian beam is two-dimensional, as in the simulations that we present in this paper, then 
\begin{equation} \label{EQ-E_y}
	E_{y}^{2D} \propto \sqrt{\frac{\sigma_0}{\sigma(x)}} \exp \left( - \frac{y^2}{\sigma^2(x)} \right),
\end{equation}
where $y$ is the non-ignorable transverse coordinate. The on-axis intensity then decreases as $\sigma_0/\sigma(x)$ away from the beam waste due to the reduced dimensionality.

\begin{figure}[tb]
\includegraphics[width=0.99\columnwidth]{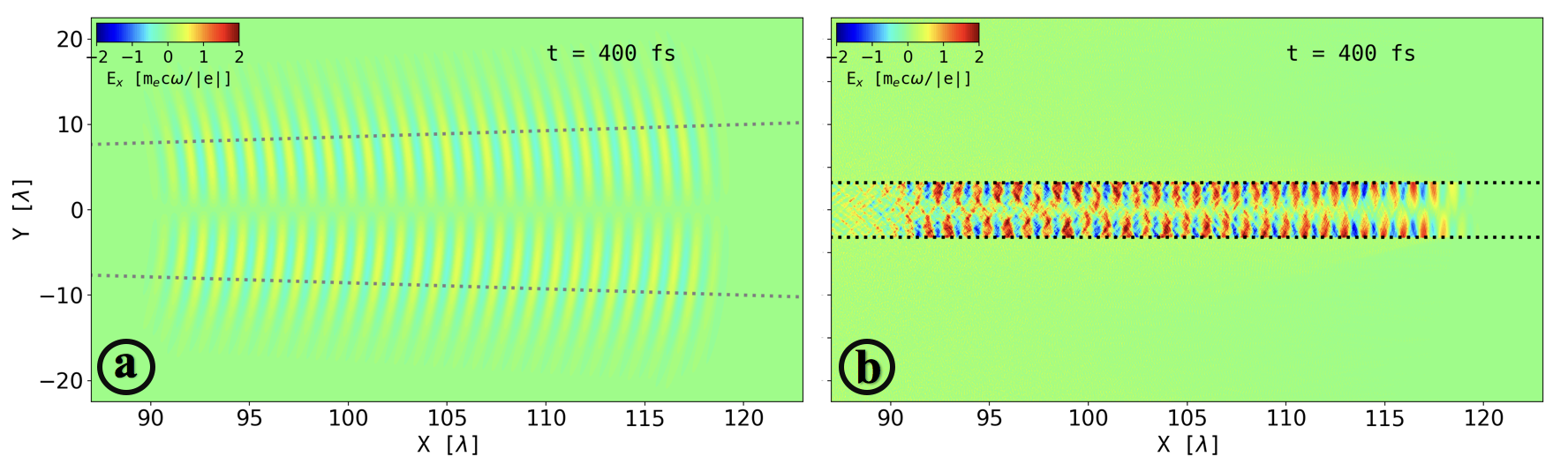}
\caption{Structure of the longitudinal electric field $E_x$ generated by a 2D Gaussian pulse in a vacuum (a) and inside a hollow-core target (b). In both cases, the focal plane is located at $x = 5$~$\mu$m, which is also the location of the target boundary in (b). The dotted curve in (a) marks the beam width $\sigma(x)$ given by Eq.~(\ref{EQ-sigma}). The dotted line in (b) marks the initial boundary of the hollow-core.} \label{fig_ex}
\end{figure}

In order to demonstrate how a hollow channel can mitigate the diffraction, we have compared two two-dimensional (2D) simulations. The `control' simulation considers only a 2D Gaussian beam, where a laser pulse with a peak intensity of $I_0 = 5.4 \times 10^{20}$ W/cm$^2$, corresponding to $a_0\approx20$, propagates in vacuum. The other beam parameters are $\sigma_0 = 4$ $\mu$m, $\lambda = 1$ $\mu$m, and $x_0=5$ $\mu$m.  The result in Fig.\ref{fig_ex}a shows that the beam width expands to $\sigma \approx 8.9$~$\mu$m by $x=100$~$\mu$m, which is in good agreement with Eq.~(\ref{EQ-sigma}). The longitudinal laser field $E_x$ must satisfy the condition $\nabla \cdot \textbf{E} = 0$, which readily gives the following estimate for its maximum amplitude: 
\begin{equation} \label{EQ-E_x}
| E_x |_{\max} \approx \frac{\lambda}{2 \pi} \left| \frac{ \partial E_y}{\partial y} \right|_{\max} \sim \frac{\lambda}{\sigma(x)} \sqrt{\frac{\sigma_0}{\sigma(x)}} E_0,
\end{equation}
where $E_0$ is amplitude of the transverse field in the focal plane located at $x = x_0$. The longitudinal field peaks away from the beam axis at $y = \sigma(x)/\sqrt{2}$, which can be shown using Eqs.~(\ref{EQ-E_y}) and (\ref{EQ-E_x}). We find that at $x=100$~$\mu$m the longitudinal field is already only $| E_x |_{\max} \approx 10^{-2} E_0$. Such a low longitudinal field is unlikely to provide efficient longitudinal acceleration.

In the second simulation, a hollow-core target with an inner radius of $r_d=3.2$~$\mu$m and an electron density of $n_e = 50 n_c$ is utilized to optically guide the laser pulse. Here $n_c \equiv m_e \omega^2 / 4 \pi e^2$ is the classical critical density that determines the density cutoff for propagation of laser pulses with $a_0 < 1$, where $\omega$ is the frequency of the laser pulse and $e$ is the electron charge. The electron density cutoff increases to roughly $a_0 n_c$ for laser pulses with $a_0 \gg 1$ due to the relativistically induced transparency. Our target remains opaque to the laser pulse, since $n_e = 50 n_c > a_0 n_c \approx 20 n_c$. The simulation result in Fig.~\ref{fig_ex}b shows the advantages of using the hollow-core channel: not only does the structured target suppress the transverse diffraction of the laser pulse, but also the laser longitudinal field is enhanced. The beam width is now limited to be less than $r_d=3.2$~$\mu$m. Already at $x = 100$~$\mu$m it is substantially smaller than the beam width in the vacuum case that is determined by the diffraction, with $\sigma \approx 9$~$\mu$m. Moreover, the transverse profile of $E_y$ is now more similar to a super-Gaussian function, with $E_y \propto \exp(-y^{\kappa} / \sigma^{\kappa})$. This leads to the longitudinal laser field being enhanced by an additional factor of $\kappa$ compared the longitudinal field in a Gaussian pulse of the same width. Figure~\ref{fig_ex}b shows that the amplitude of the longitudinal electric field in the hollow-core target is $E_x \approx 0.1 E_0$, which is ten times bigger than $E_x$ in the vacuum case at $x = 100$~$\mu$m. The observed sustained enhancement of the longitudinal laser field would play a significant role in electron acceleration.


\section{\sffamily{Quasi-static field structure inside the hollow channel}} \label{Sec-3}

In addition to the laser fields, there are also fields that are generated by the target itself in response to the laser pulse. The total force induced by these fields on the laser-accelerated electrons is dramatically different in the channel of a hollow-core target as compared to the force in a channel produced in an initially uniform target.

We first review the field structure in the ion channel that features in the `standard' approach to DLA. In an initially uniform target, the laser beam tends to expel some electrons radially outwards, creating a positively charged channel. The corresponding transverse quasi-static electric field is directed away from the axis and, in the 2D case, it is roughly given by
\begin{equation} \label{EQ-Ey_DLA}
\frac{|e| \overline{E}_y}{m_e \omega c} \approx \frac{\Delta n_i}{n_c} \frac{y \omega}{c},
\end{equation}
where $\Delta n_i$ is the uncompensated ion density that we assume to be constant across the channel. At the same time, the electrons that are present in the channel are pushed forward by the laser pulse, creating a longitudinal current. In the 2D case, the corresponding quasi-static magnetic field is directed along the $z$-axis and it is given by 
\begin{equation} \label{EQ-Bz_DLA}
\frac{|e| \overline{B}_z}{m_e \omega c}  \approx - \frac{| j_x |}{|e| n_c c}  \frac{y \omega}{c},
\end{equation}
where $j_x < 0$ is the electron current density that, for simplicity, we assume to be constant and negative (because of the forward motion) across the channel. The transverse force induced by these electric and magnetic fields on a forward moving electron in the ion channel is always directed towards the axis:
\begin{equation} \label{EQ-force_DLA}
F_y^{IC} = - |e| \left( \overline{E}_y - \frac{v_x}{c} \overline{B}_z \right) \approx - m_e \omega c \left( \frac{\Delta n_i}{n_c} + \frac{v_x}{c} \frac{| j_x |}{|e| n_c c} \right) \frac{y \omega}{c} \leq 0.
\end{equation}

\begin{figure*}[tb]
\includegraphics[width=0.9\columnwidth]{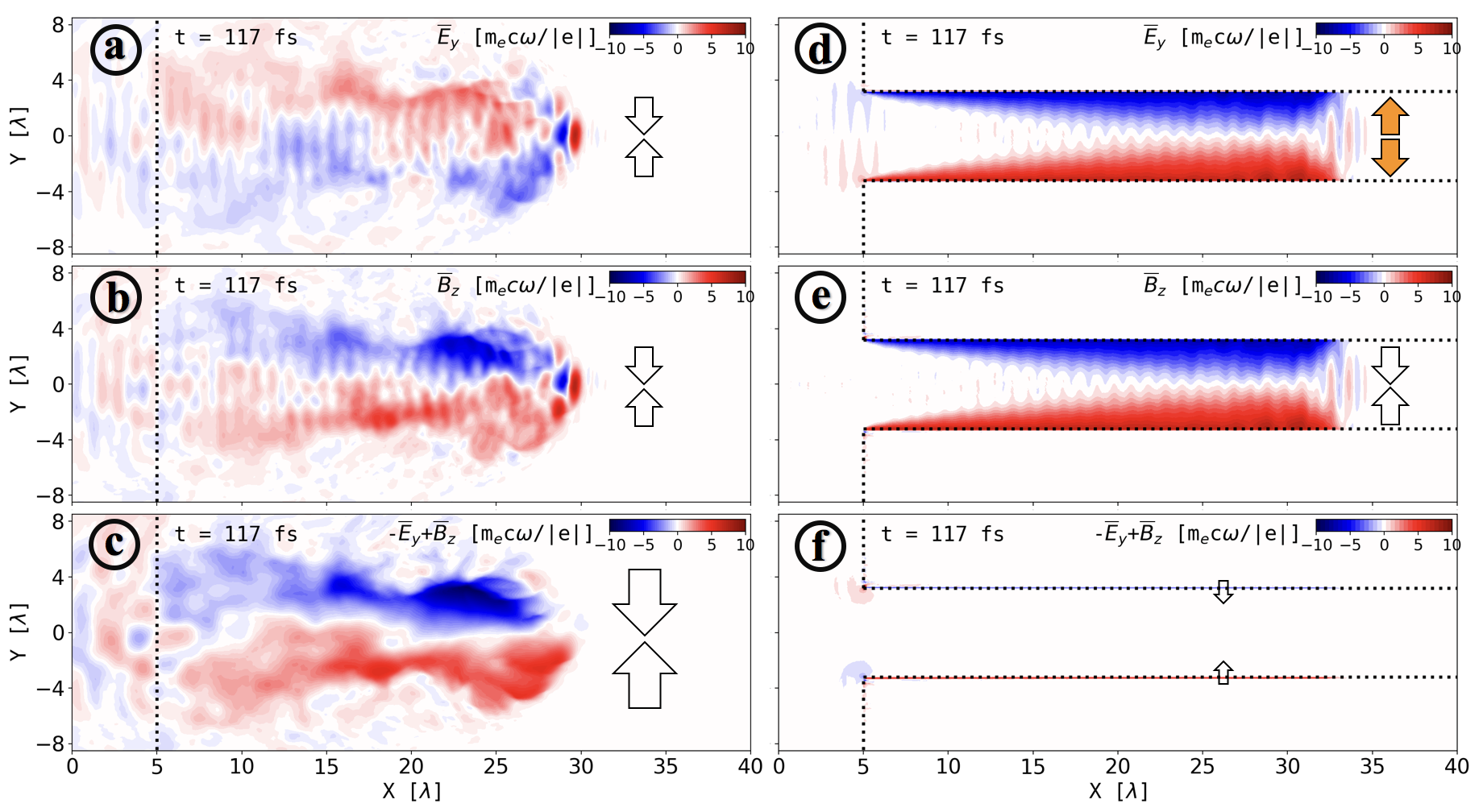}
\caption{Structure of the quasi-static electric (a and d) and magnetic fields (b and e) in an initially uniform (left) and in a hollow-core target (right). The fields are calculated by time-averaging over one laser period. The lower two panels (c and f) show the total force exerted on an ultra-relativistic forward-moving electron, $v_x \rightarrow c$. In each panel, arrows indicate the direction of a transverse force exerted on a forward-moving electron.} \label{fig_ey_bz_averaged}
\end{figure*}

Equation~(\ref{EQ-force_DLA}) illustrates a well-known result that the quasi-static plasma fields in a uniform target generate a restoring force for forward moving laser-accelerated electrons that pushes them towards the axis of the channel. We have performed a 2D PIC simulation for a uniform target with an initially uniform electron density of $n_e = 0.5 n_c$ irradiated by the laser pulse shown in Fig.~\ref{fig_ex}a. Quasi-static electric and magnetic fields are calculated by averaging the fields in the simulation over one laser period, with the corresponding time-averaged values denoted using an overhead bar. The profiles of $\overline{E}_y$ and $\overline{B}_z$, shown in Figs.~\ref{fig_ey_bz_averaged}a and \ref{fig_ey_bz_averaged}b, qualitatively agree with our estimates: $\overline{E}_y > 0$ and $\overline{B}_z < 0$ above the axis for $y>0$, whereas $\overline{E}_y < 0$ and $\overline{B}_z > 0$ below the axis for $y<0$. The combined transverse force induced by these fields and shown in Figs.~\ref{fig_ey_bz_averaged}c for an electron with $v_x \approx c$ is indeed directed towards the axis of the laser-produced channel.

In contrast to the uniform target, there are initially no particles inside the channel of the hollow-core target, which leads to a very different electric field structure. This region is populated with electrons as a result of the laser pulse interaction with the channel walls. As shown in Fig.~\ref{fig_main_structure}a, the electrons are ripped out of the wall by $E_y$ of the laser pulse. Once the injected electrons fill up the channel, the quasi-static electric field is roughly given by
\begin{equation}
\frac{|e| \overline{E}_y}{m_e \omega c} \approx -\frac{\Delta n_e}{n_c} \frac{y \omega}{c},
\end{equation}
where $n_e$ is the electron density. This field, shown in Fig.~\ref{fig_ey_bz_averaged}d, is directed away from the axis, so it is  opposite to the field of the ion channel, shown in Fig.~\ref{fig_ey_bz_averaged}a and estimated by Eq.~(\ref{EQ-Ey_DLA}). The magnetic field, however, is generated via exactly the same mechanism as in the ion channel and it can again be estimated using Eq.~(\ref{EQ-Bz_DLA}). This corresponding field structure is shown in Fig.~\ref{fig_ey_bz_averaged}e. It is convenient to express the electron current density explicitly in terms of $n_e$: $j_x = - |e| n_e u$, where $u$ is the effective velocity of the electron population in the channel. The total transverse force induced by the quasi-static electric and magnetic fields on a forward moving electron in the hollow-core target is then given by:
\begin{equation} \label{EQ-force_HC}
F_y^{HC} = - |e| \left( \overline{E}_y - \frac{v_x}{c} \overline{B}_z \right) \approx m_e \omega c \frac{n_e}{n_c} \left( 1 -  \frac{u v_x}{c^2} \right) \frac{y \omega}{c} \geq 0.
\end{equation}

A comparison of Eqs.~(\ref{EQ-force_DLA}) and (\ref{EQ-force_HC}) confirms that the change in the direction of the quasi-static electric field in the hollow-core target qualitatively alters the total transverse force exerted on laser-accelerated electrons. The total force for a forward moving electron is now expelling, $F_y^{HC} \geq 0$, instead of confining, $F_y^{IC} \leq 0$. More importantly, the forces induced by the electric and magnetic fields compensate each other with the compensation becoming significant for ultra-relativistic forward-moving electrons ($v_x \rightarrow c$) and a relativistic electron current ($u \rightarrow c$). The compensation is evident from Figs.~\ref{fig_ey_bz_averaged}c and \ref{fig_ey_bz_averaged}f, where the forces are calculated for an ultra-relativistic forward moving electron with $v_x \rightarrow c$. Even though the amplitudes of the electric, $|\overline{E}_y|$, and magnetic, $|\overline{B}_z|$, fields are higher in the hollow-core target (Figs.~\ref{fig_ey_bz_averaged}d and \ref{fig_ey_bz_averaged}e) than in the ion channel (Figs.~\ref{fig_ey_bz_averaged}a and \ref{fig_ey_bz_averaged}b), the total transverse force is drastically reduced, $|F_y^{HC}| \ll |F_y^{IC}|$.

It is important to point out that there are narrow regions at the wall of the channel where some uncompensated confining force is still present. These regions, associated with uncompensated ion charge, are aligned along the horizontal dotted lines in Fig.~\ref{fig_ey_bz_averaged}f that mark the initial boundary of the hollow-core. The force is instrumental in preventing laser-accelerated electrons from sliding out of the hollow channel. As we show in the next section of the paper, the combination of the enhanced longitudinal laser field $E_x$ and the reduction of the strong restoring force $F_y$ creates the conditions necessary for collimated longitudinal electron acceleration.



\section{\sffamily{ Longitudinal and transverse electron acceleration in the channel}} \label{Sec-4}

The transverse restoring force exerted on laser-accelerated electrons in the ion channel and shown in Fig.~\ref{fig_ey_bz_averaged}c causes transverse electron oscillations across the channel that can be successfully leveraged to enhance the electron energy gain from the laser pulse~\cite{pukhov2002strong, arefiev2012_PRL, Khudik_POP_2016}. The energetic electron bunch generated using this approach~\cite{liu2013_PRL, hu2015dense} has a considerable divergence angle that is unavoidable. In this scenario that has become synonymous with DLA, the energy is transferred from the transverse laser electric field to the transverse electron motion and only then the strong laser magnetic field converts it into the kinetic energy of the forward-directed motion. The energy transfer is then the work performed by $E_y$:
\begin{equation}
W_y = - \int |e| E_y v_y dt.
\end{equation}
There is almost no net energy gain from the quasi-static plasma field $|\overline{E}_y|$ in each transverse oscillation across the channel and the vast majority of the energy comes from the oscillating laser field. In the case of an ultra-relativistic electron, we have $E_y v_y \approx E_y c \sin \theta$, where $\theta$ is the angle between the electron velocity and the axis of the channel. Evidently, it is impossible to significantly decrease the divergence, which would be equivalent to decreasing the angle $\theta$, without effectively halting the energy transfer and thus the conventional DLA process.

Electron acceleration by longitudinal laser electric fields offers a conceptually different path for achieving large electron energies that circumvents the described difficulty. In this case, the work performed by the laser pulse goes directly towards increasing the energy of the longitudinal electron motion,
\begin{equation}
W_x = - \int |e| E_x v_x dt,
\end{equation}
without the need for the transverse oscillations to mediate the process. Even though the advantage of this approach is well recognized, there are several obstacles that have prevented its direct implementation. The approach requires a strong longitudinal laser electric field, but that can only be achieved if the laser pulse is tightly focused, as evident from Eq.~(\ref{EQ-E_x}). However, a tightly focused laser pulse has a short Rayleigh length, which limits the length over which an electron can be accelerated in a vacuum. Additionally, there is a significant transverse ponderomotive force in such a pulse, so the electron is quickly expelled radially outwards as the laser intensity ramps up. As a result, the electron might not even experience the peak intensity and there is also a possibility that it might be expelled before it reaches the focal region with a strong longitudinal electric field while moving forward.

\begin{figure}[tb]
\includegraphics[width=0.9\columnwidth]{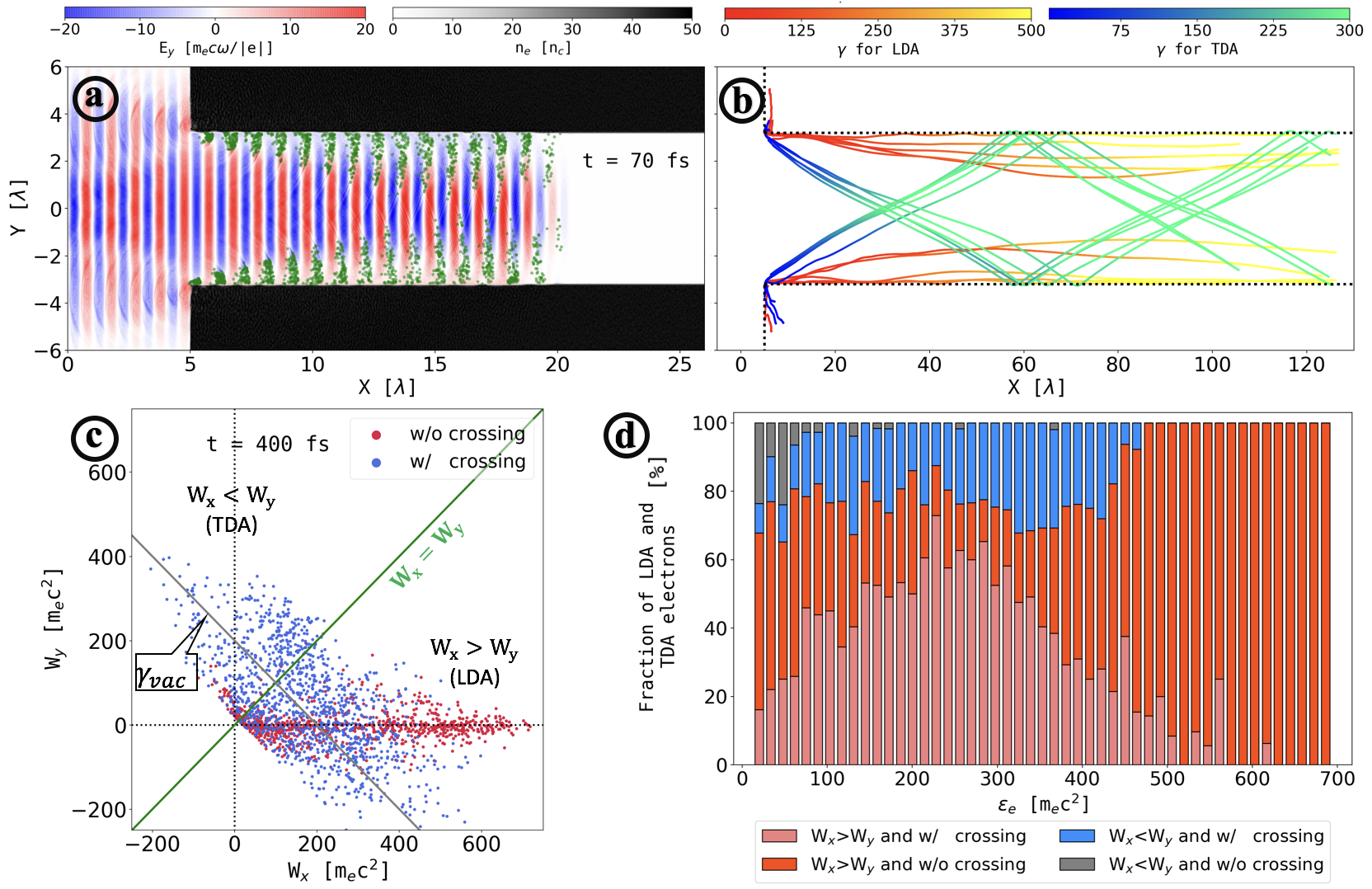}
\caption{Electron injection and acceleration in a hollow-core target. (a)
 The transverse electric field $E_y$ is plotted on top of the electron density $n_e$, while the injected electrons are shown by green dots. (b) Representative trajectories of laser-accelerated electrons are plotted over 400 fs. The $\gamma$-factor for LDA and TDA electrons is shown with the red-yellow and blue-green color-schemes. (c) The work performed by $E_x$ and $E_y$ on each of 2000 randomly selected injected electron over 400 fs is presented as a scatter-plot, where the color indicates whether the electron crossed the axis (blue) or not (red). (d) The relative fraction of LDA and TDA electrons from panel c is shown as a function of final energy (at $t = 400$~fs). Additional color-coding for each group indicates how many electrons have crossed the axis and how many have not.} \label{fig_main_structure}
\end{figure}

The hollow-core target removes the obstacles that exist in the case of pure vacuum acceleration and thus enables effective acceleration by the longitudinal electric field. We have already demonstrated in Section~\ref{Sec-2} that the longitudinal laser electric field $E_x$ can be enhanced and maintained at that level over distances that greatly exceed the Rayleigh length. The electrons accelerated inside the channel of the hollow-core target originate from the peripheral wall region. They are extracted from the wall and injected into the channel by the transverse laser electric field $E_y$ at the channel entrance, as shown in Fig.~\ref{fig_main_structure}a. It is critical that these electrons are injected immediately into a strong laser field. The transverse expelling ponderomotove force is approximately given by
\begin{equation}
F_y \approx - \frac{1}{\gamma} \frac{\partial a_0^2}{\partial y} m_e c^2 ,
\end{equation}
where $\gamma \equiv 1 / \sqrt{1 - v^2/c^2}$ is the relativistic factor.
Since $F_y \propto 1/\gamma$, its role is quickly reduced as the electrons become ultra-relativistic in the strong laser field and this is the reason why the injected electrons are not easily expelled.

The electron extraction can only take place if $E_y$ is pointing into the wall of the channel. This is why the extraction location alternates between the upper and the lower channel walls. Figure~\ref{fig_main_structure}a clearly illustrates this periodicity, with the injected electron bunches at the lower wall being shifted by half a laser cycle relative to the electron bunches injected at the upper wall. In order to generate the plot of the injected electrons, we first randomly selected 4\% of the electrons in the simulation and then those with $\gamma > 1.5$ and $|y| < 3.2$ $\mu$m were plotted with green dots. The charge and current of the injected electrons sustain the channel electric and magnetic fields shown in Figs.~\ref{fig_ey_bz_averaged}d and \ref{fig_ey_bz_averaged}e. It is worth pointing out that these are ultra-relativistic electrons with a predominantly longitudinal velocity, so the effective velocity $u$ that determines the electron current is very close to the speed of light. As discussed in Sec.~\ref{Sec-3}, this leads to the significant reduction of the transverse force given by Eq.~(\ref{EQ-force_HC}) and shown in Fig.~\ref{fig_ey_bz_averaged}f.

Our focus is the role of the longitudinal acceleration in achieving high electron energies, so we divide electrons based on the amount of work performed by longitudinal and transverse electric fields into two groups to aid our analysis: 
\begin{itemize}
\item Group \#1 with $W_x > W_y$: The longitudinal field $E_x$ does more work on these electrons than the transverse field $E_y$. For compactness, we call this group LDA, which stands for longitudinal dominant acceleration.
\item Group \#2 with $W_x \leq W_y$: The work by the longitudinal field $E_x$ does not exceed that of the transverse field $E_y$. For compactness, we call this group TDA, which stands for transverse dominant acceleration.
\end{itemize}
Figure~\ref{fig_main_structure}c shows how the total work is partitioned between $W_x$ and $W_y$ for the injected electrons at $t = 400$ fs, where each dot represents a macro-particle. The dots below the solid green line are LDA electrons. The plot was generated by randomly selecting 2000 injected electrons with $|y| < 3.2$ $\mu$m.

Figure~\ref{fig_main_structure}d shows the relative fraction of LDA electrons as a function of energy. A significant portion of the electrons in Fig.~\ref{fig_main_structure}d are LDA electrons regardless of the energy, which confirms that the longitudinal laser electric field indeed plays a major role in electron acceleration. It is remarkable that an appreciable fraction of the injected electrons are very energetic at this stage, as their energy exceeds the so-called vacuum limit $\varepsilon_{vac}$ for an electron in a plane wave. These are the electrons above the gray line in Fig.~\ref{fig_main_structure}c. The vacuum limit is the maximum energy for an initially immobile electron in a plane electromagnetic wave with peak amplitude of $a_0$: $\varepsilon_{vac}  \equiv \gamma_{vac} m_e c^2$, where $\gamma_{vac} \equiv 1 + a_0^2/2$. The observed energy increase is a clear advantage of using the hollow-core target instead of employing a pure vacuum setup.

Tracking of the injected electrons has revealed a significant difference in trajectories between TDA and LDA electrons. A vast majority of TDA electrons perform transverse oscillations across the channel that are characterized by crossing its axis ($y = 0$). Figure~\ref{fig_main_structure}d quantifies the fraction of the TDA electrons that have crossed the axis as a function of the electron energy. Examples of these trajectories are shown in Fig.~\ref{fig_main_structure}b, where the blue-green color-scheme indicates the relativistic $\gamma$-factor along the trajectory. In contrast to the TDA electrons, the majority of the energetic LDA electrons never cross the axis of the channel. These are the electrons with energies above $500 m_e c^2$ in Fig.~\ref{fig_main_structure}d. The energetic LDA electrons remain relatively close to the wall. The corresponding trajectories are shown in Fig.~\ref{fig_main_structure}b using the red-yellow color-scheme, where the color indicates the relativistic $\gamma$-factor along the trajectory.

\begin{figure}[tb]
\includegraphics[width=0.9\columnwidth]{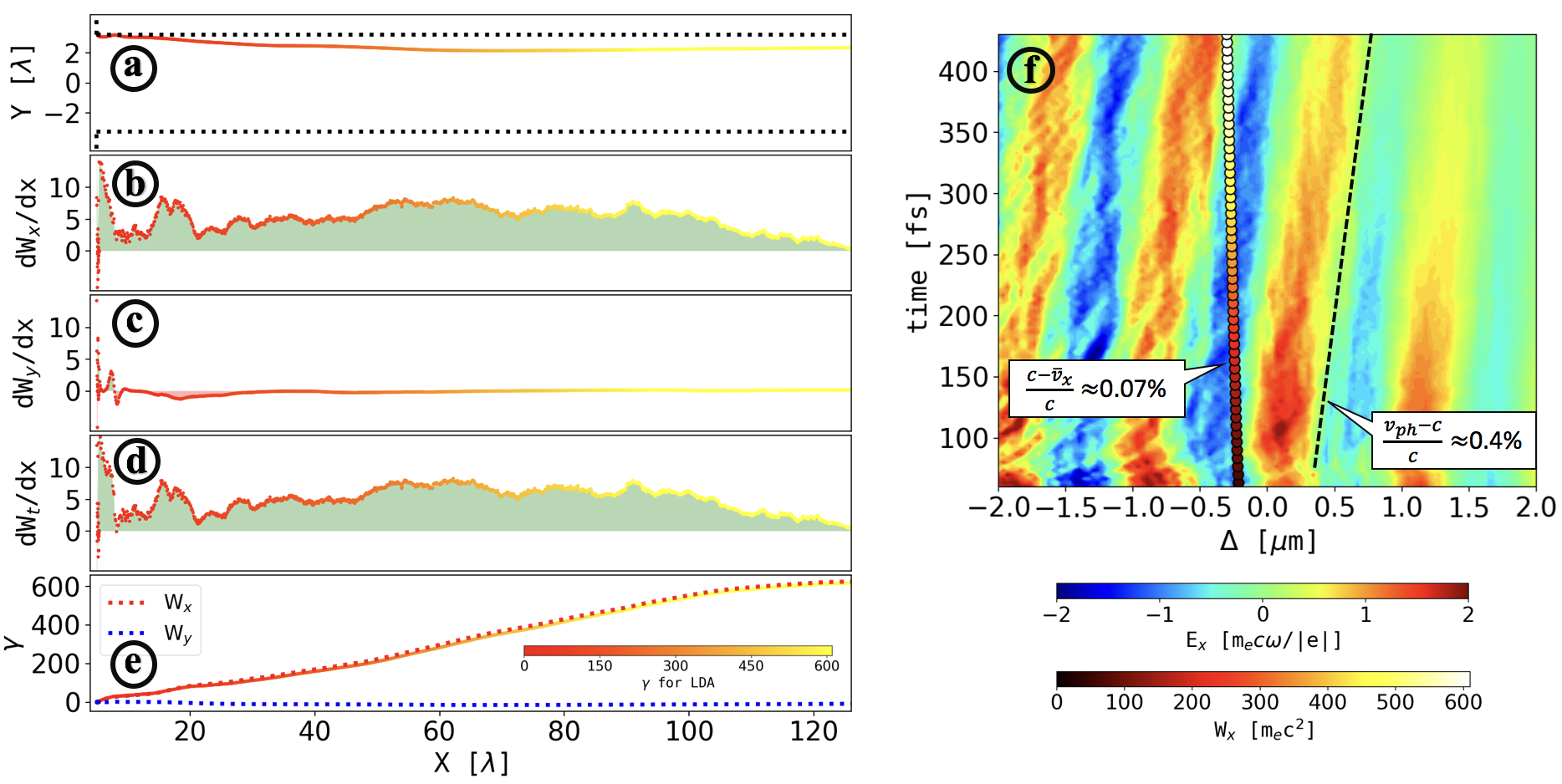}
\caption{Detailed tracking information for a representative energetic LDA electron. The color-coding along the curves in panels (a) - (e) is the relativistic $\gamma$-factor. The quantities $d W_{x,y,t}/dx$ are normalized to $m_e c^2 /\lambda$. The electron trajectory in panel (f) is plotted on top of the longitudinal electric field in a window moving forward with a speed of light, $\Delta \equiv x-ct-12\lambda$. The color-coding of the markers indicates the work $W_x$ performed by $E_x$.} \label{fig_power_x}
\end{figure}

The described qualitative differences in the trajectories are directly related to the key features in the two acceleration mechanisms. In what follows, we elucidate this aspect by tracking two representative electrons: one energetic LDA electron with $\max \gamma \approx 600$ that never crosses the axis of the channel and one energetic TDA electron with $\max \gamma \approx 400$ that performs a full oscillation across the channel.

Figure~\ref{fig_power_x} provides detailed information about a selected LDA electron. As evident from the electron trajectory in Fig.~\ref{fig_power_x}a, the longitudinal motion in the forward direction dominates. It is instructive to examine how the work performed by $E_x$ and $E_y$ changes as the electron moves along the channel. In order to do that, we plot $dW_x / dx$ (Fig.~\ref{fig_power_x}b) and $dW_y/dx$ (Fig.~\ref{fig_power_x}c) instead of plotting the accumulated quantities $W_x$ and $W_y$. Figure~\ref{fig_power_x}d shows the change in the total work, $W_t \equiv W_x + W_y$. The accumulated area under this curve gives the relativistic $\gamma$-factor (or the electron energy) plotted in Fig.~\ref{fig_power_x}e.

There are two aspects that are evident from Figs.~\ref{fig_power_x}e and \ref{fig_power_x}b: the electron energy comes predominantly from the work done by $E_x$, whereas the contribution of $E_y$ is negligible; the electron continuously gains energy from $E_x$ as it travels more than a hundred wavelengths along the channel. The influence of $E_y$ is reduced because the transverse component of the electron velocity is relatively small compared to the longitudinal component. The continuous positive work by $E_x$ suggests that the electron is moving sufficiently fast to remain in the accelerating part of the wave. This is further confirmed by Fig.~\ref{fig_power_x}f, where $E_x$ and the electron longitudinal location are plotted in a window moving with the speed of light, such that $\Delta \equiv x-ct-12\lambda$. The electron indeed remains in the region with negative $E_x$ over hundreds of fs, which allows it to efficiently gain energy.

Clearly, the key to achieving LDA is to ensure that an electron remains locked in an accelerating phase of the negative longitudinal electric field for an extended period of time. The reduction of the transverse force $F_y^{HC}$ that we demonstrated in the previous section is critical: the transverse velocity remains small during the acceleration process, with the angle $\theta$ between \textbf{v} and the $x$-axis less than $2^\circ$. As a result, the average longitudinal electron velocity $\overline{v}_x$ is less than $c$ by only 0.07\%. In fact, the difference between the phase velocity of the wave, $v_{ph}$, and the speed of light is greater than $c - \overline{v}_x$. Indeed, we find that $c - \overline{v}_x \approx 7 \times 10^{-4} c$, whereas $v_{ph} - c \approx 4 \times 10^{-3} c$. This means that the energy gain in our case is limited by the super-luminosity of the wave-fronts and not by the transverse or longitudinal electron motion.

\begin{figure}[tb]
\includegraphics[width=0.9\columnwidth]{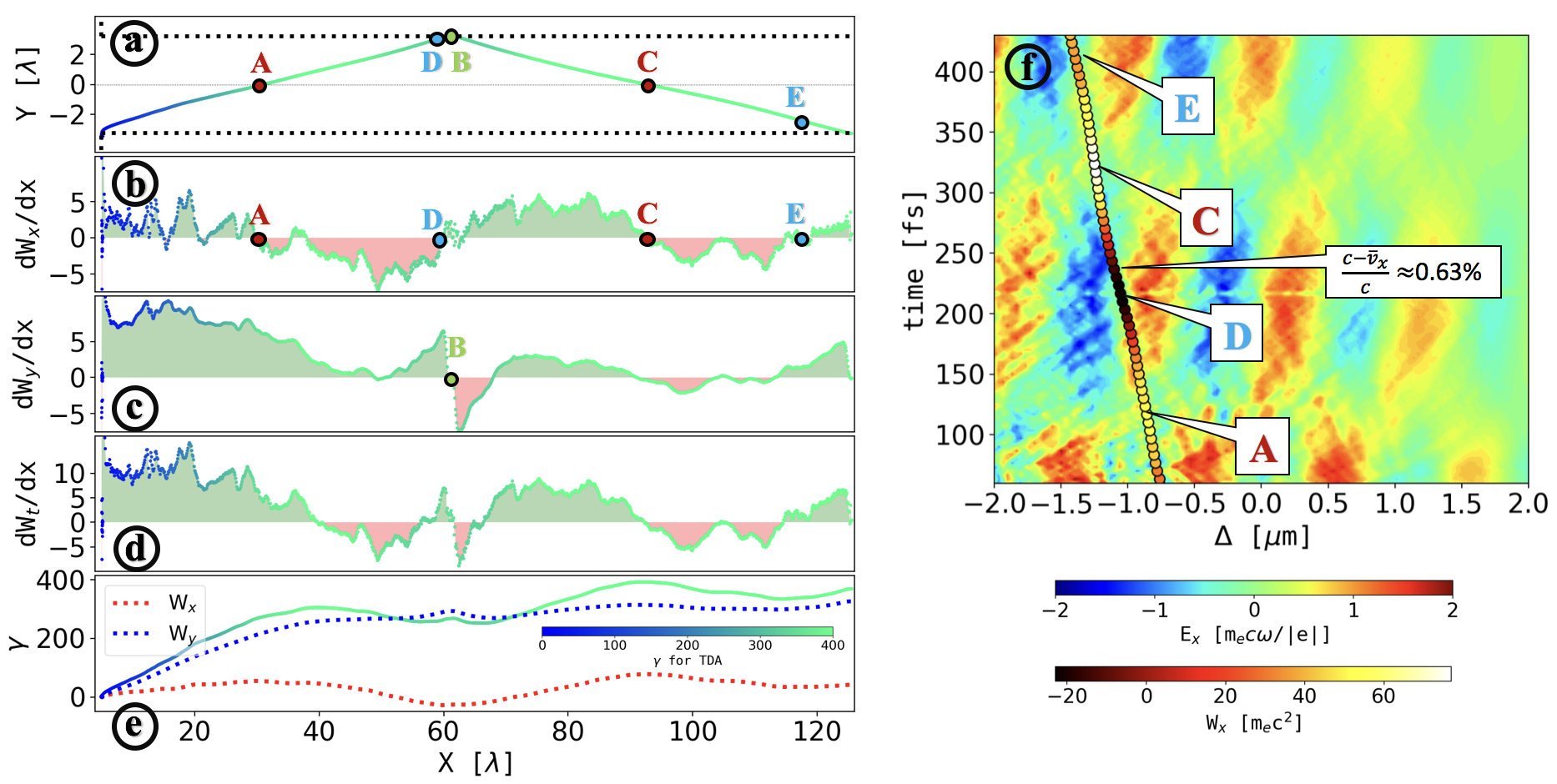}
\caption{Detailed tracking information for a representative energetic TDA electron. The color-coding along the curves in panels (a) - (e) is the relativistic $\gamma$-factor. The quantities $d W_{x,y,t}/dx$ are normalized to $m_e c^2 /\lambda $. The electron trajectory in panel (f) is plotted on top of the longitudinal electric field in a window moving forward with a speed of light, $\Delta \equiv x-ct-12\lambda$. The color-coding of the markers indicates the work $W_x$ performed by $E_x$.} \label{fig_power_y}
\end{figure}

In contrast to the considered LDA electron, a selected TDA electron whose detailed tracking information is shown in Fig.~\ref{fig_power_y} has a relatively large initial transverse velocity $v_y$ that was acquired at the injection stage. In this case, the angle between \textbf{v} and the $x$-axis is $\theta \approx 7^\circ$. This allows the electron to traverse the channel while moving forward, as can be seen in Fig.~\ref{fig_power_y}a, which, in turn, impacts the work done by $E_x$. The longitudinal laser electric field $E_x$ is essentially an odd function of $y$, i.e. $E_x (y) \approx -E_x (-y)$. This means that the electron switches between accelerating and decelerating regions by crossing the axis of the channel even if it has a sufficiently high longitudinal velocity to remain in the accelerating phase of the laser field for an extended period of time. This effect is well-illustrated in Figs.~\ref{fig_power_y}a and \ref{fig_power_y}b, where A and C mark the crossings of the channel axis. The work by $E_x$ changes its sign not only at A and C, but also at D and E where the electron slides out of the decelerating or accelerating phase due to its relative longitudinal motion with respect to the laser wave-fronts. The frequent alternations between the positive and negative work performed by $E_x$ significantly reduce the electron energy gain from longitudinal laser electric field.

Most of the energy for the considered TDA electron comes from the transverse laser electric field (see Fig.~\ref{fig_power_y}e), which is the conventional DLA mechanism. Electron reflections off the channel walls (a reflection point is marked with a B) can aid the energy exchange between the electron and the transverse laser electric field $E_y$ by changing the direction of $v_y$. As $v_y$ flips, the compensation that usually exists between energy gain and loss in a purely vacuum case is disrupted. This can potentially lead to an appreciable energy increase, but this approach requires multiple bounces across the channel. It is worth pointing out that $c - \overline{v}_x \approx 6 \times 10^{-3} c$ is comparable to $v_{ph} - c \approx 4 \times 10^{-3} c$, so the super-luminosity is not a major factor limiting the energy gain for the considered electron.


\begin{figure}[tb]
\includegraphics[width=0.9\columnwidth]{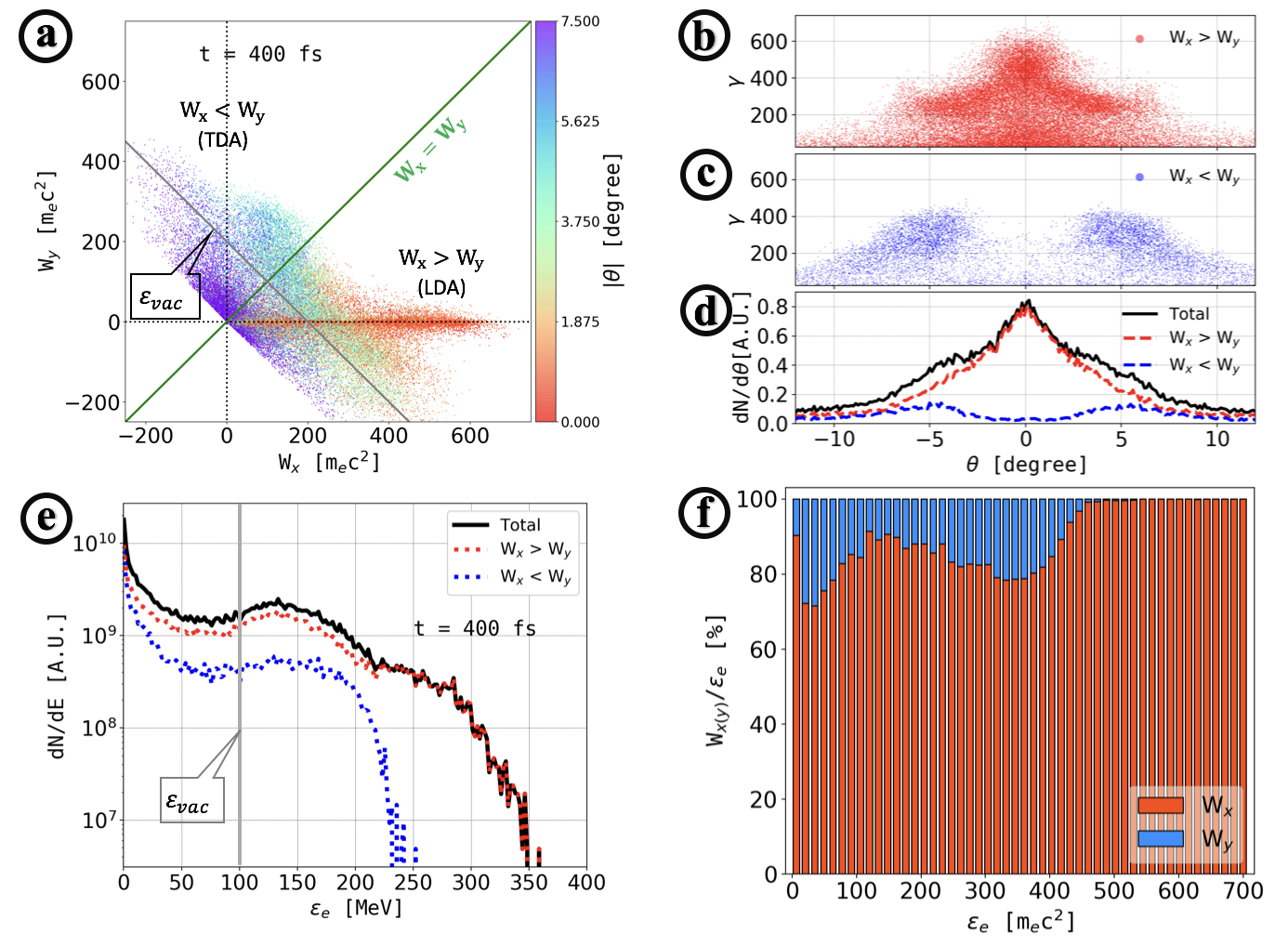}
\caption{Angular distribution of accelerated electrons and their energy spectra. (a) The work performed by $E_x$ and $E_y$ on injected electrons over 400 fs is presented as a scatter-plot, with the color indicating the angle between the electron momentum and the $x$-axis. (b) and (c) LDA and TDA electrons are presented as scatter-plots in the $(\theta, \gamma)$-space. (d) The angular distributions represent the scatter-plots shown in panels b and c. (e) Energy spectra of injected electrons are shown after 400 fs. The spectra of LDA and TDA electrons are shown in red and blue. (f) The relative fraction of the energy contributed by $E_x$ and $E_y$ is shown as a function of the electron energy $\varepsilon_e$. These values are averaged over all injected electrons in a given energy bin.} \label{fig_angle}
\end{figure}


\section{\sffamily{Collimated energetic electrons}} \label{Sec-5}

In Section~\ref{Sec-4}, we examined how transverse and longitudinal laser electric fields contribute to acceleration of the injected electrons in the channel of a hollow-core target. In what follows, we examine the angular distribution of these electrons and how the two groups of electrons, LDA and TDA, contribute to it.

Figure~\ref{fig_angle}a is the same scatter-plot of the accelerated electrons in the $(W_x, W_y)$-space as Fig.~\ref{fig_main_structure}c, but the color-coding now indicates the polar angle $\theta$ between the electron momentum and the axis of the channel. In order to generate this plot, we selected all macro-particles that are inside the channel, $|y| < 3.2$~$\mu$m, after 400 fs into the simulation. The angle is then calculated by computing $\theta = \arctan(p_y/p_x)$ for each macroparticle, where $p_x$ and $p_y$ are the momentum components provided by the simulation. Electrons with a small angle $\theta$ are tightly clustered around $W_y = 0$ in Fig.~\ref{fig_angle}a, which suggests that that the longitudinal acceleration (LDA) tends to produce well-collimated energetic electrons, but the transverse acceleration (TDA) does not. 

As evident from Fig.~\ref{fig_angle}a, energetic TDA electrons have a polar angle whose absolute value exceeds $3^\circ$, while some fraction of the highly collimated energetic LDA electrons have a polar angle $|\theta| < 1^\circ$. In order to gain more insight about the angular distribution of the energetic electrons, we plotted them in the $(\theta, \gamma)$-space, with the scatter-plots for LDA and TDA electrons shown in Figs.~\ref{fig_angle}b and \ref{fig_angle}c. The curves in Fig.~\ref{fig_angle}d show the total angular electron distribution and the corresponding contributions from LDA and TDA electrons that represent the scatter-plots from Figs.~\ref{fig_angle}b and \ref{fig_angle}c. The angular distribution of the TDA electrons has two symmetric peaks at $\theta \approx \pm 5^\circ$. In contrast to that, the angular distribution of LDA electrons has a single sharp peak at $\theta \approx 0^\circ$ with a full width half maximum of approximately $4^\circ$.

The difference in the angular spread between the TDA and LDA electrons is determined by the transverse electron velocity gained during their injection into the channel. However, a detailed study of the injection dynamics and the factors determining the injection velocity goes beyond the scope of the current work. It is worth pointing out that although the angular spread of the TDA electrons is worse than that of the LDA electrons, it is still significantly better than the angular spread during the conventional DLA regime because of the reduced net restoring force inside the hollow channel. 

The energy spectra of LDA and TDA electrons at $t = 400$~fs are shown in Fig.~\ref{fig_angle}e. The LDA electrons have a much higher cutoff energy (350 MeV) than the TDA electrons (250 MeV), but both energies significantly exceed $\varepsilon_{vac} \approx 100$ MeV. The so-called super-ponderomotive part of the spectrum ($\varepsilon_e > \varepsilon_{vac}$) is rather flat as compared to an exponentially decaying spectrum generated in a uniform target with the help of the conventional DLA (for example, see Fig.~5 in Ref.~\cite{arefiev2016beyond}). The longitudinal laser electric field $E_x$ does most of the work over the entire range of electron energies, as can be seen in Fig.~\ref{fig_angle}f. This figure provides a relative fraction of the energy contributed by $E_x$ and $E_y$, i.e. $W_x/\varepsilon_e$ and $W_y/\varepsilon_e$, for electrons in a given energy bin.


Even though the emphasis has been on the longitudinal laser electric field, both components of the laser field are critical for producing energetic electrons with $\varepsilon_e > \varepsilon_{vac}$ in a hollow-core target. We demonstrate this by performing an additional simulation where the polarization of the laser electric field is rotated by $\pi/2$ in the $(y,z)$-plane. In this case, there is no transverse laser electric field that is normal to the channel boundary. As a consequence, the electron injection into the channel is severely suppressed. Moreover, there is no strong longitudinal laser electric field, because the condition $\nabla \cdot \textbf{E} = 0$ can be satisfied for the transverse component $E_z$ without any significant $E_x$. As a result, the maximum electron energy in this case is around 4 MeV.


\section{\sffamily{Summary and discussion}} \label{Sec-6}

We have presented a new mechanism for generating collimated beams of energetic electrons that utilizes longitudinal laser electric fields. By focusing a relativistically intense laser pulse into a tailored hollow-core target, the electrons are injected into the hollow region and undergo acceleration dominated by the longitudinal electric field. The hollow-core target is essential for enhancing the longitudinal fields and maintaining them over a distance much longer than the Rayleigh length by guiding the laser pulse.

The substantial angular divergence of electron beams produced by direct laser acceleration is often considered as an inherent negative feature of the mechanism. The divergence however arises primarily because the standard approach relies on transverse electron oscillations and their interplay with the transverse electric fields of the laser pulse. Our approach is conceptually different, as it directly leverages longitudinal laser electric fields that are present in a tightly focused laser beam. Transverse oscillations that mediate the energy transfer between the laser pulse and the electrons in the traditional approach are no longer required.

In our scheme, transverse electron oscillations are reduced by creating a negatively charged electron channel, as opposed to a positively charged channel that is essential for the traditional DLA approach. Electrons are injected by the transverse laser electric field into the channel and then they are accelerated forward by the pulse, creating an electron current. The forces from electric and magnetic fields of this electron population compensate each other, creating a favorable configuration without a strong restoring force.

Our 2D PIC simulations confirm that a highly directed electron beam with a low angular spread is produced due to a combination of the enhancement of the longitudinal laser field and the elimination of transverse oscillations. The electron cut-off energy of 350 MeV is much greater than the vacuum limit of $\varepsilon_{vac} = (1 + a_0^2/2)m_e c^2 \approx 100$ MeV. Meanwhile, unlike the conventional DLA electron beam with a typical opening angle of $20^\circ$, our electron beam has an opening angle of less than 5$^\circ$.  

The key feature of our regime is that most of the energy (across the entire spectrum) is transferred to the electrons by the longitudinal laser electric field and, given a sufficient acceleration distance, super-ponderomotive energies can be realized without sacrificing the collimation. We speculatively suggest that this mechanism may have played a role in previous studies focused on high-energy laser-driven photon sources~\cite{yi2016bright,yu2018generation}, although further work is required to determine if such a connection exists. It also remains to be understood if the same mechanism can be realized by utilizing a microwire array~\cite{jiang2014effects, jiang2016microengineering}. Previous results ~\cite{jiang2014effects, jiang2016microengineering} showed a well-collimated electron beam, bu no super-ponderomotive electrons. We speculate that this might be a consequence of a relatively short acceleration length allowed to the electrons.


\section{\sffamily{Acknowledgements}}

The work was supported by the National Basic Research Program of China (Grant No.2013CBA01502), NSFC (Grant Nos.11535001), National Grand Instrument Project (2012YQ030142), and the National Science Foundation (Grant No. 1632777). Simulations were performed using the EPOCH code (developed under UK EPSRC Grants No. EP/G054940/1, No. EP/G055165/1, and No. EP/ G056803/1) using HPC resources provided by the TACC at the University of Texas.


\appendix

\section*{\sffamily{Appendix: PIC simulation parameters}} \label{PIC_parameters}

All simulations presented in this work are two-dimensional. They were carried out using a fully relativistic particle-in-cell code EPOCH~\cite{arber2015contemporary}. For all but one simulation, the simulation box is a rectangle in the $(x,y)$-plane that is $130 \lambda$ long ($0 \leq x \leq 130 \lambda$) and $24 \lambda$ wide ($-12 \lambda \leq y \leq 12 \lambda$). Here our grid size is $\Delta x = 1/50 \lambda$ along the $x$-axis and $\Delta y = 1/20 \lambda$ along the $y$-axis. In the simulation shown in Fig.~\ref{fig_ex}, we used a wider box that is $48 \lambda$ to perform the comparison of laser diffraction between the vacuum and the channel cases (the resolution is the same).

The laser pulse enters the domain at $x = 0$. It is a Gaussian beam in $y$ with an axis at $y=0$ and a focal plane at $x=5$ $\mu$m. The laser wavelength is $\lambda = 1$ $\mu$m. In the vacuum case, the peak intensity is $I_0 = 5.4 \times 10^{20}$ W/cm$^2$, corresponding to $a_0\approx20$. The pulse has a trapezoid temporal profile with 8.33 fs up/down ramps (2.5 laser periods each) and an 83.3 fs flattop (25 laser periods). We use open boundary conditions for the fields~\cite{arber2015contemporary}.

The structured target is initialized as a fully ionized uniform plasma slab (5$\mu$m $\leq x \leq $ 130$\mu$m) with a hollow core ($|y| \leq 3.2$ $\mu$m). The slab electron density is $n_e = 50 n_c$ and it is initialized using 50 macro-particles per cell representing electrons. The ions are fully ionized carbon ions. Initially, their density is equal to $n_i = n_e / 6$ and it is initialized using 25 macro-particles per cell representing ions. The ions are treated as immobile in our simulations to clearly distinguish the effect of the electron dynamics. The simulation with a uniform target uses a similar setup. In this case, the electron density is $n_e = 0.5 n_c$, there is no channel, and the ion density is initialized using 10 macro-articles per cell. 

It has been previously shown that the temporal resolution is critical for obtaining the correct electron energies during simulations of the direct laser acceleration~\cite{arefiev2015temporal}. The criterion can be reformulated as $\Delta x \leq \lambda/a_0$ for $\Delta t \approx \Delta x /c$. To ensure the validity of our results, an extra simulation was performed with $\Delta x = 1/100 \lambda$ and $\Delta y = 1/50 \lambda$. The resulting electron energy spectra and electron angular divergence showed no significant change compared to those discussed in the manuscript. In order to aid the particle tracking process by reducing the data set, we used the resolution with $\Delta x = 1/50 \lambda$ and $\Delta y = 1/20 \lambda$.


\bibliographystyle{ieeetr}
\bibliography{aa}
\end{document}